\documentclass[prl,twocolumn,showpacs]{revtex4}
\usepackage[dvips]{graphicx}
\usepackage{latexsym}
\usepackage{bm}
\begin{document}
\newcommand{\fig}[2]{\includegraphics[width=#1]{#2}}
\newcommand{\pprl}{Phys. Rev. Lett. \ } 
\newcommand{\pprb}{Phys. Rev. {B}} 
\newcommand{\be}{\begin{equation}}
\newcommand{\ee}{\end{equation}}
\newcommand{\bea}{\begin{eqnarray}} 
\newcommand{\eea}{\end{eqnarray}}
\newcommand{\nn}{\nonumber} 
\newcommand{\tjv}{{$t$-$J$-$V$}}
\newcommand{\tj}{$t$-$J$}
\newcommand{\sbar}{{\bar\sigma}}
\newcommand{\la}{\langle}
\newcommand{\ra}{\rangle} 
\newcommand{\dg}{\dagger}
\newcommand{\br}{{\bf{r}}} 
\newcommand{\tnh}{{\rm tanh}}
\newcommand{\sh}{{\rm sech}} 
\newcommand{\pll}{\parallel} 
\newcommand{\upa}{\uparrow} 
\newcommand{\dna}{\downarrow}
\newcommand{\lw}{\Lambda_\omega}
\newcommand{\lwp}{\Lambda_\omega^\prime}
\newcommand{\tphi}{{\tilde\phi}}
\newcommand{\ru}{{U}}
\newcommand{\rj}{{J}}
\newcommand{\rt}{{t}}
\newcommand{\rru}{{U_r}}
\newcommand{\rrj}{{J_r}}
\newcommand{\rrt}{{t_r}}

\title{
Pseudogap, competing order and coexistence of
staggered flux and $d$-wave pairing in high-temperature superconductors
}

\author{Sen Zhou and Ziqiang Wang}
\affiliation{Department of Physics, Boston College, Chestnut Hill, MA 02467}

\date{January 22, 2004}

\begin{abstract}

We study the \tjv model of a
doped Mott insulator in connection to high-T$_c$ superconductors.
The nearest neighbor Coulomb interaction ($V$) is
treated quantum mechanically on equal footing
as the antiferromagnetic exchange interaction ($J$).
Motivated by the $SU(2)$ symmetry at half-filling, we construct
a large-N theory which allows a systematic study of the interplay 
between staggered flux order and superconductivity upon doping.
We solve the model in the large-N limit and obtain the ground state
properties and the phase diagram as a function of doping.
We discuss the competition and the coexistence of the staggered flux 
and the d-wave superconductivity in the underdoped regime and
the disappearance of superconductivity in the overdoped regime.

\typeout{polish abstract}
\end{abstract}

\pacs{74.25.Jb, 71.10.Fd, 74.20.Mn, 71.10.Hf}

\maketitle

The physics of doped Mott insulators has played
a central role in the study of high-T$_c$ superconductors 
\cite{anderson,pwanew}. Traditionally, the Mott insulating state is derived 
from the strong {\it onsite} Coulomb repulsion and naturally accounts for
the antiferromagnetic (AF) ordered state at half-filling \cite{anderson}.
There is increasing evidence that,
due to large quantum fluctuations in spin-1/2 systems,
several exotic quantum spin liquid states
are likely hidden under the AF Mott insulating ground state.
These states are competitive in energy and form a
basis for the resonating valence bond
route to superconductivity \cite{anderson}. Several important, 
yet unresolved problems remain: Is the d-wave superconductor (dSC) the 
ground state when the long-range AF order is
destroyed by the coherent motion of the doped holes? 
Does the competition between the dSC and other forms of order account 
for the pseudogap phenomenon in the underdoped regime? 
Does the physics of doped Mott insulator explain the disappearance of 
superconductivity in the overdoped regime?

In this paper we demonstrate that {\it long-range} Coulomb interactions,
not crucial for discussing the Mott insulating state, play
an important role in addressing these issues associated with
doped Mott insulators. Because charge fluctuations are limited by 
the density of {\it doped} holes, long-range Coulomb interactions have
been thought to be not important near the stoichiometric Mott limit. 
However, since the strength of the interaction is large 
compared to the Fermi energy, its effects become quickly significant 
with doping. We consider the microscopic
\tjv model where $V$ stands for the 
nearest-neighbor (NN) Coulomb repulsion. 
As in the \tj model,
%The infinite {\it on-site} repulsion,
the constraint of no double occupancy leads to an
insulating state at half-filling.
It is well known that the d-wave pairing state and the staggered
flux phase (SFP) \cite{marston,kotliar,tdkotliar}, identical at half-filling
owing to an $SU(2)$ gauge symmetry \cite{affleck}, are two
competing phases at low doping.
The SFP has a plaquette flux density wave (orbital
current) order with $d$-wave symmetry.
It is a metallic state with hole pockets and a pseudogap of the
same symmetry as the dSC. Thus the SFP serves as a 
candidate pseudogap phase for underdoped cuprates, 
much as in the
%the weak-coupling 
$d$-density wave scenario \cite{ddw}. 
However, in the standard \tj model, the ground state becomes
the dSC upon doping \cite{kotliarliu}. The SFP was found to be unstable to 
superconductivity \cite{zhang,wangkotliar} and only exists at finite 
temperatures in a very narrow regime near half-filling \cite{ubbenslee}.

We show that the NN Coulomb interaction introduces competing
and coexisting order and changes significantly the structure of
the phase diagram. 
It is crucial to treat the $V$-term quantum mechanically, 
on equal footing as the exchange $J$-term, since both pairing and screening 
originate at the same energy scale. 
Motivated by the $SU(2)$ symmetry at half-filling,
we construct an $SU(N,q)$ ($q=$quaternion) generalization of
the \tjv model.
%in the slave boson formulation.
While the $SU(2N)$ theory \cite{sun} favors the normal
state and the $SP(2N)$ theory \cite{spn} favors
the superconducting (SC) state in the large-N limit, the $SU(N,q)$ theory
accommodates both staggered flux (SF) and SC order, and
allows a systematic study of their interplay.
In the large-N limit, 
we find that the SF order coexists with the
dSC below a critical doping $x_c$ where a quantum
phase transition to a pure dSC phase takes place. 
The SC phase is destroyed in the overdoped regime 
beyond $x_p > x_c$. At finite temperatures, the onset of the SF order 
is at $T_{SF}>T_c$, giving rise to a pseudogap SFP 
above the SC transition temperature $T_c$. 
We discuss the phase diagram, the signatures of the 
coexistence and the pseudogap phase.

We start with the large-N generalization of the 
\tjv model on a square lattice, $H=H_{tJ}+H_V$,
\bea
H_{tJ}&=&-{t\over N}\sum_{\langle i,j\rangle}
\left(c_{i,\sigma}^\dagger c_{j\sigma}
+{\rm h.c.}\right) + {J\over N}\sum_{\langle i,j\rangle}
{\bf S_i}\cdot {\bf S_j}
\label{htj}\\
H_V&=&{V\over N^3}\sum_{\langle i,j\rangle}n_i n_j, \quad V=V_c-{1\over4}J.
\label{hv}
\eea
The operator $c_{i\sigma}^\dagger$ creates an electron with spin 
$\sigma=1,{\bar1};\cdots; N,{\bar N}$ 
at site $i$ under the constraint of no double occupancy for each
spin flavor, $n_i=c_{i\sigma}^\dagger c_{i\sigma}\leq N$.
The NN Coulomb repulsion between
the electrons is given by $V_c$, which is grouped
together with $-(J/4)n_in_j$ from
the exchange term and represented by the $V$-term
in Eq.~(\ref{hv}). The usual \tjv model corresponds to $N=1$.
The projected Hilbert space can be treated in the slave boson
formalism by writing $c_{i\sigma}^\dagger =f_{i\sigma}^\dagger b_i$,
where $f_{i\sigma}^\dagger$ is a spin-carrying fermion and
$b_i$  a spinless boson keeping track of empty sites, with the constraint
$
f_{i\sigma}^\dagger f_{i\sigma}+ b_i^\dagger b_i=N.
%\label{constraints}
$

Let us introduce the pseudo-spinors, 
$\Psi_i=(\psi_{i\uparrow},\psi_{i\downarrow})$,
%which transform under the fundamentals of $SU(N,q)$,
\be
\psi_{i\alpha\uparrow}=\left(\begin{array}{c}
f_{i\alpha} \\
f_{i\bar\alpha}^\dagger
\end{array}\right), \quad
\alpha=1,\cdots,N, 
\label{spinor}
\ee
and $\bar\psi_{i\downarrow}={\bf C}{\psi}_{i\uparrow}$ where
${\bf C}=i\sigma_y\otimes{\bf 1}$.
Here ${\bf 1}_{\alpha\beta}$ is the identity matrix in the
flavor space and $(\sigma_y)_{\tau\tau^\prime}$ is the Pauli matrix 
acting on the spins. Under a unitary transformation $g_i$,
$\psi_{i\uparrow}\to g_i\psi_{i\uparrow}$ and
$\psi_{i\downarrow}^\dagger \to \psi_{i\downarrow}^\dagger g_i^\dagger$,
provided that $g_i^t{\bf C}g_i={\bf C}$, i.e. $g_i\in SU(N,q)$.
The $SU(N,q)$ fermion bilinear
$
U_{ij}=\Psi_i\Psi_j^\dagger
$
transforms as $U_{ij}\to g_iU_{ij}g_j^\dagger$.
To construct the large-N theory, we form the
flavor singlet by tracing over the flavor indices,
\be
u_{ij}={\rm Tr}U_{ij}
=\left(\begin{array}{cc}
-{\hat\chi}_{ij}^* &{\hat\Delta}_{ij} \\
{\hat\Delta}_{ij}^* & {\hat\chi}_{ij}
\end{array} \right),
\label{uij}
\ee
which is a {\it quaternion} with elements:
$\hat\chi_{ij}= \sum_\alpha(f_{i\alpha}^\dagger f_{j\alpha}
+f_{i\bar\alpha}^\dagger f_{j\bar\alpha})$ and
$\hat\Delta_{ij}= \sum_\alpha(f_{i\alpha}^\dagger f_{j\bar\alpha}^\dagger
-f_{i\bar\alpha}^\dagger f_{j\alpha}^\dagger)$. They
represent the bond (SF) and pairing degrees of freedom.
In terms of $u$, the Hamiltonian can be written as,
\bea
H_{tJ}=&-&{t\over2N}\sum_{\langle i,j\rangle}\left[b_i b_j^\dagger
\sum_{\tau}\psi_{i\tau}^\dagger (1+\sigma_z)\psi_{j\tau}
+{\rm h.c.}\right] 
\label{htj1} \\
&-&{J\over8N}\sum_{\langle i,j\rangle}{\rm tr} (u_{ij}^\dagger u_{ij})
+\sum_{i,\tau}\psi_{i\tau}^\dagger (i\lambda_i-\mu_f)\sigma_z\psi_{i\tau}.
\nonumber \\
H_V=&-&{V\over2N^3}\sum_{\langle i,j\rangle}b_i^\dagger b_i b_j^\dagger b_j
{\rm tr}( u_{ij}^\dagger\sigma_z u_{ij} \sigma_z).
\label{hv1}
\eea
Here ${\rm tr}$ stands for the trace in spinor space, $\lambda_i$ is the
Lagrange multiplier enforcing the local constraints, and $\mu_f$ fixes the
average fermion density. 
Note that for general $N$, $H_V$ cannot
be written solely in terms of four fermions using the constraints.

In the large-N limit, the bosons condense at low temperatures
$\langle b_i\rangle=\sqrt{N x_i}$, where $x_i$ is the local doping
concentration \cite{wang}. Note that since
charge fluctuations are suppressed, the $V$-term
vanishes with vanishing boson amplitude on approaching half-filling
where the $SU(2)$ symmetry is restored. 
The interactions are decoupled by introducing the saddle point values 
$\chi_{ij}=\langle \hat\chi_{ij}\rangle/N$ and 
$\Delta_{ij}=\langle\hat\Delta_{ij}\rangle/N$.
The saddle-point Hamiltonian becomes
$H=\sum_{\langle i,j\rangle}H_{ij}-\sum_i(\mu_f-i\lambda_i)\hat\chi_{ii}$
\bea
H_{ij}=&-&t\sqrt{x_i x_j}\hat\chi_{ij}
-g_\chi\chi_{ij}^\dagger \hat\chi_{ij}
-g_\Delta\Delta_{ij}^\dagger \hat\Delta_{ij}+{\rm h.c.}
\nonumber \\
&+&N( g_\chi\vert\chi_{ij}\vert^2+g_\Delta \vert\Delta_{ij}\vert^2),
\label{mf}
\eea
where $g_\chi=J/4+x_ix_j V$ and $g_\Delta=J/4-x_ix_j V$ are the fermion 
coupling constants to the bond and paring order parameters respectively.
The effect of $V$ is to
reduce pairing while enhance the SF ordering. 
We focus on the uniform solutions with $x_i=x$, $i\lambda_i=\lambda$,
$\chi_{ij}=\chi_1+i(-1)^{i_x+j_y}\chi_2$ and
$\Delta_{ij}=\Delta_s+i(-1)^{i_y+j_y}\Delta_d$
where $\chi_{1,2}$ and $\Delta_{s,d}$ are real. The flux per plaquette is
given by $\phi=4\tan^{-1}(\chi_2/\chi_1)$. 
$\Delta_s$ and $\Delta_d$ are the $s$-wave and $d$-wave pairing order
parameters, respectively. 
The free energy per site is obtained analytically,
\bea
F=&-&2TN{1\over N_s}\sum_{k,s=\pm}\ln\left[2\cosh(E_{ks}/2T)\right]
\nonumber \\
&+&2N(g_\Delta\vert\Delta\vert^2+g_\chi\vert\chi\vert^2)-N \mu x,
\label{f}
\eea
where $\mu=\mu_f-\lambda$. The $k$-sum is over the reduced zone
due to the doubling of the unit cell, and $E_{ks}$ describe
the two branches of the quasiparticle (QP) dispersion
$$
E_{ks}=\left[\bigl(\sqrt{\alpha_k^2+\beta_k^2}+s\mu\bigr)^2
+4g_\Delta^2(\gamma_+^2\Delta_s^2+\gamma_-^2\Delta_d^2)\right]^{1/2}
%\label{ek}
$$
with $\gamma_\pm=\cos k_x \pm \cos k_y$, $\alpha_k=2
(xt+g_\chi \chi_1)\gamma_+$, and $\beta_k=2g_\chi\chi_2 \gamma_-$.
Minimizing the free energy in Eq.~(\ref{f}), we obtain the
self-consistent equations,
\bea
\Delta_s &=& {{\Delta_s g_{\Delta}}\over N_s}{\sum_{ks}}\tanh
\left( {E_{ks} \over {2T}}\right) {\gamma_+^2 \over E_{ks}},\nonumber\\
\Delta_d &=&{{\Delta_d g_{\Delta}}\over N_s}{\sum_{ks}}\tanh
\left( {E_{ks} \over {2T}}\right) {\gamma_-^2 \over E_{ks}},\nonumber\\
\chi_1 &=&{1 \over {2N_s}}{\sum_{ks}}\tanh\left( {E_{ks} \over {2T}}
\right) {{\alpha_k \gamma_+} \over E_{ks}}\left[ 1+{{s\mu} \over
{(\alpha_k^2 +\beta_k^2 )^{1/2}}}\right],
\nonumber \\
\chi_2 &=&{1\over 2N_s}{\sum_{ks}}\tanh
\left( {E_{ks}\over{2T}}\right) {{\beta_k\gamma_-} \over E_{ks}}\left[1
+{{s\mu}\over{(\alpha_k^2 +\beta_k^2 )^{1/2}}}\right],
\nonumber\\
x &=&-{1\over N_s}{\sum_{ks}}\tanh\left( {E_{ks}\over{2T}}\right)
{{s(\alpha_k^2 +\beta_k^2 )^{1/2}+\mu} \over E_{ks}}.
\label{mfp}
\eea
In the absence of the NN Coulomb interaction, $V=0$ and
$g_\chi=g_\Delta$. The equations for $\chi_2$ and $\Delta_d$ imply
that they cannot be simultaneously nonzero except at half-filling ($\mu=0$) 
\cite{zhang}.
The NN Coulomb interaction $V>0$ leads to $g_\chi>g_\Delta$
away from half-filling, making it possible for the coexistence of the two 
types of order.

We solve the self-consistent equations in (\ref{mfp}) using $J$ as the energy 
unit and $t=2J$, $V=2J$. The $s$-wave pairing 
$\Delta_s$ is found to be zero. In Fig.~1, $\Delta_d$, $\chi_1$
and $\chi_2$ are plotted at $T=0$ as a function of doping $x$.
The results for $V=0$ is also plotted for comparison.
The NN Coulomb interaction introduces two important features.
(i) The dSC order develops upon doping, but is destroyed
beyond a critical doping $x_p\simeq0.26$, which is
smaller than the naive estimate from the condition $g_\Delta=0$.
The suppression of pairing by $V$ in the overdoped regime is consistent
with recent quantum Monte Carlo studies of
the $t$-$U$-$V$ model \cite{sorella}. 
(ii) The SF order develops
below a critical doping $x_c\simeq0.1$ and coexists with the dSC.
There is a reduction of $\Delta_d$ concomitant with the development
of the SF ($\chi_2$) below $x_c$, indicative of competing order.
The values of $x_c$ and $x_p$ depend on the band structure. 
A next NN hopping $t^\prime<0$ moves both to higher values.
Although the $d$-wave pairing amplitude $\Delta_d$
extrapolates to a nonzero value at half-filling,
the two states are identical by the $SU(2)$ gauge symmetry.
\begin{figure}
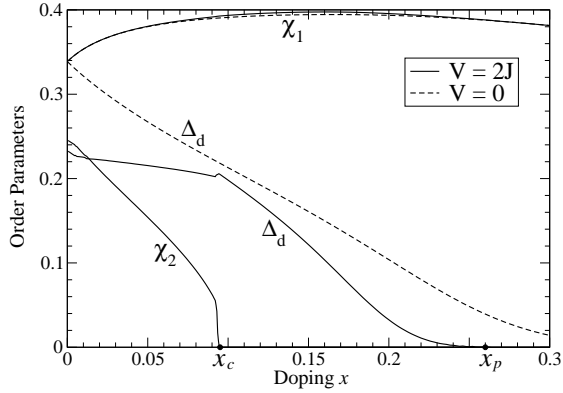

\begin{center}
%\vskip-2.8cm
%\hspace*{-3.8cm}
\fig{2.9in}{fig1.eps}
\vskip-2mm
\caption{ 
The order parameters versus doping at $T=0$.
%for $V=0$ (dashed lines) and $V=2J$ (solid lines). 
%Inset: the ground state energy
%difference between the coexistence phase and the pure dSC at $V=2J$.
}
\label{fig:fig1}
\vskip-8mm
\end{center}
\end{figure}

In Fig.~2, the phase diagram of the \tjv model is shown in the
plane of temperature $T$ versus doping $x$. The phase boundaries are
the temperatures $T_{SF}$ and $T_c$ at which
the SF and dSC order parameters vanish, respectively. The topology of
the phase diagram resembles the one proposed for the cuprate high-$T_c$
superconductors by invoking competing order and 
quantum criticality \cite{ddw}. The dSC sets in below $x_p$. 
The transition temperature $T_c$ rises with decreasing 
doping and reaches its optimal value somewhat below the quantum 
critical point (QCP) at $x_c$ where coexisting SF order emerges. 
The $T_{SF}$ line
shows a bulge when crossing the $T_c$ line at $x_c^\prime\simeq0.12$, 
making it possible for a thermal transition from the dSC to the coexistence 
phase for $x_c<x<x_c^\prime$. The
phase boundary rises almost vertically at low temperatures, and the
transition is weakly first order as is evidenced
by the small discontinuity in the order parameters (Fig.~1).
For $x<x_c^\prime$,  $T_c< T_{SF}$,
giving rise to a pseudogap SFP with orbital currents 
for $T_c<T<T_{SF}$ in the underdoped regime.
The phase diagram and the nature of the pseudogap phase are different 
from those in the $SU(2)$ formulation of the \tj model \cite{leewen}. 
In that theory, the $SU(2)$ symmetry at half-filling is extended to 
finite doping via the introduction of a slave boson doublet. The pseudogap
phase is a fluctuating state between gauge equivalent SFP and the dSC.
The static SF order develops only inside vortices where SC order 
is suppressed by the external magnetic field \cite{leewenvortex}.
The present theory explores the consequences of symmetry breaking
by doping, through the NN Coulomb interaction in particular, which
stabilizes the static order of the SF in the pseudogap phase.
\begin{figure}
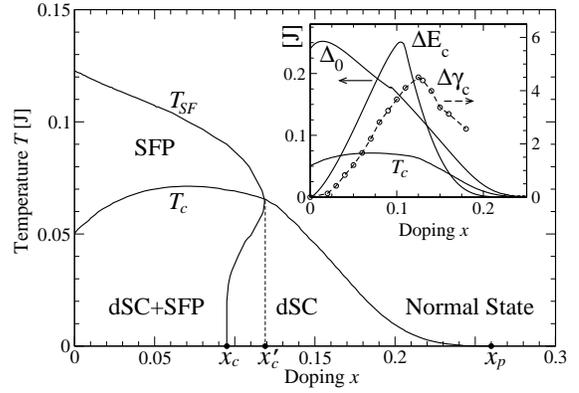

\begin{center}
%\vskip-2.8cm
%\hspace*{-3.8cm}
\fig{2.9in}{fig2.eps}
\vskip-2mm
\caption{The large-N phase diagram of the \tjv model showing a QCP at
$x_c$, a pure dSC, a coexistence phase of 
dSC and SF, and a finite temperature pseudogap SF phase.
Inset: $T=0$ single-particle gap $\Delta_0$, 
condensation energy $\Delta E_c(\times50)$, 
and specific heat coefficient jump $\Delta\gamma_c$ versus doping.
}
\label{fig:fig2}
\vskip-8mm
\end{center}
\end{figure}

To illustrate the electronic structures, we calculate the tunneling 
density of states (LDOS) measured by STM
\be
N(\omega)=-{\rm Re}\int_0^\infty dt e^{i\omega t}\theta(t)
\langle\{c_{i\sigma}^\dagger(0),c_{i\sigma}(t)\}\rangle.
\label{ldos}
\ee
In Fig.~3, we show the typical LDOS and the
QP dispersion. Fig.~3(a) corresponds to the pure dSC
at $x=0.14$ and $T=0$. The LDOS shows $d$-wave gap features. The two peaks 
near the gap edges come from the large state density 
at the points marked as $X$ at $(\pi,0)$ and $P$ at the nearby minimum
on the dispersion, and their symmetry equivalent points.
Note that the distance between neither set of peaks
equals to $2\Delta$ where $\Delta=4g_\Delta\Delta_d$ is the
maximum of the $d$-wave gap function. 
One must thus be careful with extracting the SC gap
from the peak-to-peak distance even with high STM resolution.
The LDOS in the coexistence phase is shown in Fig.~3(c) 
at $x=0.06$ and $T=0$. Compared to Fig.~3(a) in the dSC,  
there is an emergent in-gap resonance peak (marked A) 
particle-hole reflected on both sides of zero-bias. It arises from the 
resonant scattering of the QP from state ${\bf k}$ to 
state ${\bf k}+(\pi,\pi)$ due to period doubling.
A similar situation arises in the SF state around a superconducting
vortex in the SU(2) formulation of the t-J model \cite{kishine}.
As shown in Fig.~3(d) for the $s=+$ branch,
at the resonant energy, the distorted ellipses touch at point A,
the low-energy minimum in the dispersion plotted in Fig.~3(c).
These resonance peaks, albeit small in weight, are
ubiquitous signatures of coexisting order and should be detectable
by high resolution STM. Near the gap edges, there are four sets
of peaks (two are prominent) due to the doubling of the unit cell. 
They are labeled by $X_\pm$ and $P_\pm$ and marked on the 
QP dispersion. It is instructive to compare these to
the LDOS in the pseudogap SFP at $x=0.06$ and $T=0.08J$ shown in Fig.~3(b).
The low energy resonance peaks disappear and the LDOS 
shifts to the unoccupied side. Interestingly, the minimum of the LDOS in 
the SFP coincides with the M point next to the in-gap resonance
in the coexistence phase. The resonance energy and the spectral shift
depend on the band structure and doping and are
on the order of $0.043J\sim6$meV for the parameters used in Fig.~3.
\begin{figure}
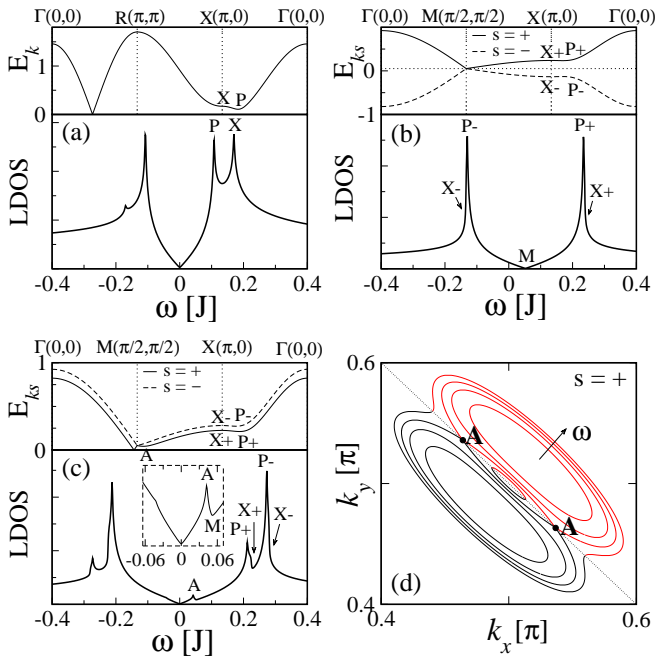

\begin{center}
%\vskip-2.8cm
%\hspace*{-3.8cm}
\fig{3.4in}{fig3.eps}
\vskip-2mm
\caption{
The LDOS and the QP dispersion in dSC (a),
SF phase (b), coexistence phase of SF and dSC (c).
(d) Equipotential contours at $\omega=0.03, 0.04, 0.043, 0.05J$,
showing resonant scattering at $\omega_A=0.043J$.
}
\label{fig:fig3}
\vskip-8mm
\end{center}
\end{figure}

An important consequence of the competing order is 
the emergence of two energy gaps which can be,
however, difficult to observe at low temperatures \cite{tallon}. 
In the inset of Fig.~2, the gap to single-particle excitations $\Delta_0$ 
obtained from the LDOS at $T=0$ is plotted versus doping,
which qualitatively agrees with the low temperature ARPES data.
It interpolates between a geometric average of the gaps associated
with the SF and SC order for $x<x_c$ and the gap of the pure dSC
for $x>x_c$ and closes at $x_p$.
The only sign of competing order in $\Delta_0$ is the
mild cusp-like singularity near $x_c$. 
Signatures of competing order are more apparent in thermodynamic
quantities \cite{ddw,tallon,wuliu}. In the same figure, the jump 
$\Delta\gamma_c$ in the specific heat coefficient $\gamma(T)=C_p(T)/T$ 
at $T_c$ and the condensation energy per site $\Delta E_c$ at $T=0$ are
shown as a function of doping. $\Delta\gamma_c$ rises with decreasing
doping on the overdoped side. 
With the opening of the pseudogap due to SF order, the entropy in the 
normal state is reduced, $\Delta\gamma_c$ turns around at $x_c^\prime$
and drops continuously to small values in the underdoped
regime, in qualitative agreement with specific heat measurements 
\cite{tallon}. The condensation energy $\Delta E_c$ grows almost
linearly with doping in the underdoped regime and peaks around
$x_c$. The closeness in the free energy between the SFP and dSC,
the smallness of the condensation energy,
together with a small superfluid density that scales with 
$x$ promote the appearance of cheap vortices
in the underdoped regime as emphasized in the $SU(2)$ theory 
\cite{leewenvortex}, except that static orbital currents in
the present theory live in the bulk as well as inside the vortex.

In summary, we have presented a microscopic large-N theory for
the competition between the exchange and the finite-range Coulomb 
interactions in doped Mott insulators. The \tjv 
model shows competing and coexisting order of SF and dSC which can be viewed
as a product of $SU(2)$ symmetry breaking by doping through the NN 
Coulomb interaction. The phase diagram exhibits a QCP separating the 
coexistence phase from a pure dSC, a finite temperature pseudogap SFP 
in the underdoped regime, and complete suppression of superconductivity beyond
an overdoping. At very low doping, the phase diagram does not
match that of the cuprates. The discrepancy is
most obvious in the nonzero $T_c$ that extends to infinitesimal doping, whereas
the SC state in the cuprates disappears at a finite doping below which
a spin glass phase prevails. The lack of destruction
of the dSC at low doping is because the renormalized NN Coulomb interaction 
must scale to zero 
at half-filling due to the constraint forbidding charge fluctuations. 
We expect the effects of doping induced inhomogeneity
\cite{pan} due to poor screening of the dopant ionic
potential to become very important in this regime \cite{wang}.
As a result, the local doping concentration (LDC) fluctuates
dramatically and its interplay with AF may lead
to the destruction of the SC state and the emergence
of spin glass and insulating behaviors. 
Furthermore, the onset of SF order will
fluctuate spatially with the LDC, smearing out the sharp 
finite temperature transition to the pseudogap SFP.

This work is supported in part by DOE Grants Nos. DE-FG02-99ER45747,
DE-FG02-02ER63404, and ACS Grant No. 39498-AC5M.

\end{document}